 \documentclass[final,5p,times,twocolumn,authoryear]{elsarticle}

\usepackage{amssymb}
\usepackage{amsmath}

\newcommand{\lineWL}[2]{#1$\lambda$#2~\AA}

\newcommand{\Ha}{$\text{H}_\alpha$}

\newcommand{\HaWL}{\lineWL{\Ha}{6563}}

\newcommand{\Alfven}{Alfv\'{e}n}

\journal{High Energy Astrophysics}

\begin{document}

\begin{frontmatter}



\title{{\it Guitar}  Nebula: extreme accelerator in extreme environment}


\author[inasan,kfu]{Nikonorov Igor Nikolaevich}
\ead{inikonorov@inasan.ru}
\author[inasan]{Barkov Maxim Vladimirovich}
\author[purdue]{Lyutikov Maxim}

\affiliation[inasan]{organization={Department of stellar physics and evolution, Institute of Astronomy of the Russian Academy of Sciences},
            addressline={48 Pyatnitskaya str.}, 
            city={Moscow},
            postcode={119017}, 
            state={Moscow},
            country={Russia}}

\affiliation[kfu]{organization={``Cosmic X-ray sources'' Research Laboratory, Kazan Federal University},
            addressline={18 Kremlyovskaya str.}, 
            city={Kazan},
            postcode={420008}, 
            state={Republic of Tatarstan},
            country={Russia}}

\affiliation[purdue]{organization={Department of Physics and Astronomy, Purdue University},
            addressline={525 Northwestern Avenue}, 
            city={West Lafayette},
            postcode={47907-2036}, 
            state={IN},
            country={USA}}

\begin{abstract}
{\it Guitar} nebula is a prime example of a class of bow-shock pulsar wind nebulae (PWNe), powered by a wind of a supersonically moving neutron star. Bow-shock PWNe can probe particle  acceleration processes in relativistic pulsar winds, as well as the structure of the  interstellar medium (ISM). We demonstrate that the {\it Guitar}  is an exceptional object in  a number of ways. First, particles escaping the PWN and forming the X-ray ``kinetic jet'' need to be accelerated to the energies corresponding to  the maximal electric potential of the neutron star  $\eta_\text{acc}\gtrsim 3/4$ : it is another example of the class of extreme accelerators. Second, exceptionally bright \Ha{} emission requires that the central pulsar PSR J2225+6535 passes through a dense, low ionization ISM region. Bright X-ray emission of the ``kinetic jet'' then also requires exceptionally high magnetic field, $\sim 100~\mu$G. We hypothesize that {\it Guitar}  passes through the one of long-predicted, narrow dense shells of an old supernova remnant, currently in the ``pressure-driven snowplow'' regime.
\end{abstract}



\begin{keyword}
supernova remnants \sep galactic cosmic rays \sep pulsars \sep interstellar medium



\end{keyword}

\end{frontmatter}




\section{Introduction \label{intro}}

Pulsars produce ultrarelativistic winds; their interaction with surrounding media forms Pulsar Wind  Nebulae (PWNe),  which  shine  from radio to hard gamma-rays band \citep{2004vhec.book.....A,2006ARA&A..44...17G}.

A particular type of PWNe is formed when a fast moving  pulsar escapes the preceding supernova ejecta. Interaction of the pulsar wind with the interstellar medium (ISM) then forms  bow-shock PWNe \citep{2019MNRAS.484.4760B,2019MNRAS.488.5690O}. Bow-shock PWNe (BSPWNe) can be used to probe the structure of ISM on scales down to the the stand-off distance (a distance between a pulsar and a bow shock apex, $\sim 10^{16}$~cm, Equation~(\ref{eq:Rs})) \citep{2025PASA...42...79N, 2020MNRAS.497.2605B,2019MNRAS.485.2041B}.

In modeling BSPWNe, various, seemingly unrelated phenomena should be taken into account: overall morphology (including possible density variations in the external medium), internal geometry of the pulsar  (relative  direction of the spin and velocity, magnetic inclination angle - bullet, frisbee and cartwheel geometries, magnetic inclination angle and magnetic hoop stresses in the pulsar wind) \citep{2019MNRAS.484.4760B}, particle acceleration in PWNe, kinetic jets produced by BSPWNe \citep{2008A&A...490L...3B,2019MNRAS.485.2041B,2019MNRAS.489L..28B}, and, specifically, dynamic (time and spatially dependent)  \Ha{} emission \citep{2025PASA...42...79N}. The resulting self-consistent model should allow us to assess parameters at the bow-shock apex and the  properties of pulsar wind.

In this paper,  we argue that {\it Guitar} PWN, associated with PSR J2225+6535, is exceptional in not one, but two ways: (i)  it accelerates particles up to the theoretical maximum (it is an example of an {\it extreme accelerator}); (ii) it currently  passes through a highly unusual ISM phase (high total density, low ionization, high magnetic field),  produced by a previous supernova explosion, unrelated to the pulsar powering the {\it Guitar}.

The plan of the work is as follows. In  \S \ref{sec:BSPWNe_description} we review the structure of BSPWN, and expected high-energy and \Ha{} signatures and outline connections between them. In \S \ref{sec:Guitar_exceptionality} we show, that all of the observed PWNe, except {\it Guitar}, fall in line with expectations: {\it Guitar} Nebula stands out both in X-rays and in \Ha{} emission. In \S \ref{sec:scenario} we offer a plausible explanation (that the central pulsar passes through a special layer in the ISM, formed by an old supernova remnant (SNR) in the snowplow regime), and explain the tension between expected and observed dispersion measure of pulsar with our scenario and give predictions on the rotation measure. \S \ref{sec:conclusions} concludes the manuscript.

\section{Bow-shock PWNe: morphology, \Ha{}, high energy emission and kinetic jets \label{sec:BSPWNe_description}}

Consider a pulsar with spin-down power $\dot{E}$ moving with velocity $V_\text{NS}$ through an external medium of the number-density $n_\text{ISM}$ and mean molecular mass $\mu_\text{ISM}$. Initially partially ionized ISM becomes fully ionized in the head part of the flow; the resulting stand-off distance ($r_s$) is \citep{1993Natur.362..133C}:
\begin{align}
r_s  &= \sqrt{ \frac{\dot{E}}{4 \pi \mu_\text{ISM} m_p n_\text{ISM} V_\text{NS}^2 c} } = \notag \\ 
     &= 1.3\times 10^{15} \dot{E}_{33.1}^{1/2} \mu_{\text{ISM}, 0}^{-1/2} n_{\text{ISM}, 0}^{-1/2} V_{\text{NS},7.9}^{-1} \mbox{ cm.} \label{eq:Rs}
\end{align}
Hereafter, $m_p$ is a proton mass and the normalization is $A_x = A \times 10^{-x}$ CGS -- Gaussian units.

Equation~(\ref{eq:Rs}) assume highly supersonic motion with pulsar Mach number ($M_\text{NS} =  V_\text{NS}/c_\text{ISM} \gg 1$, $c_\text{ISM}$ is the speed of sound in the ISM) and is subject to a number of modifications. 

As for the  external medium, the dominant  variability comes from the density structure \citep{2019MNRAS.484.4760B,2019MNRAS.484.1475T}. Another more subtle ISM factor affecting BSPWN morphology is the ionization structure of the ISM and the associated mass-loading effects within the PWN \citep{2003MNRAS.339..623L,2015MNRAS.454.3886M,2018MNRAS.481.3394O}.

Important data comes from the emission in \HaWL{} spectral line, as BSPWNe were systematically observed only in this particular  optical line. The largest survey encompasses 9 objects \citep{2014ApJ...784..154B}. Spectra of the head  part of the  ``{\it Guitar}'' show  other hydrogen Balmer series spectral lines as well \citep{1993Natur.362..133C, 2013ffep.confE..67D} (another similar case is a PWN embedded in CTB 80 SNR, observed in multiple forbidden lines \citep{1988ApJ...331L.121H}).

In the previous work \citep[][ hereafter NBL25]{2025PASA...42...79N} we have built the procedure for \Ha{} luminosity calibration of isolated BSPWNe in order to compare observed objects with our hydrodynamic models with detailed treatment of gas ionization state. In the approximation of ideal hydrodynamics and weak external magnetic field the expected \Ha{} luminosity relative to the spin-down power is then (for $0.1 \lesssim V_{\text{NS},8} \lesssim 2$):
\begin{equation}
        \frac{L_{\text{H}_\alpha}}{\dot{E}} \left( V_\text{NS} \right)_\text{NBL25}
          = \left( 1.75 V_{\text{NS}, 8}^{-2} + 4.00 V_{\text{NS}, 8}^{-1} - 2.61 \right) \times 10^{-6}. \label{eq:expectedHa}
\end{equation}

Dynamic effects of the external magnetic field  on the structure of bow-shock PWNe are likely to be negligible (although see Section~\ref{sec:Bism_bowshock_config}). In a magnetized medium the important velocity is not the speed of sound, but the fast velocity:
\begin{equation}
v_f^2 = c_s^2 + v_A^2 = c_s^2 \left(1+\frac{1}{\beta_\text{ISM}} \right),
\end{equation}
where $v_A$ is \Alfven{} velocity and $\beta_\text{ISM} = c_s^2/v_A^2$ is the plasma beta-parameter. By conventional wisdom, $\beta_\text{ISM} \sim 1-100 \geq 1$. Typical values are listed in Table~\ref{tab:ISM_typical_quantities}. As we argue below (Section~\ref{sec:scenario}), there are regions in ISM with $\beta _\text{ISM} < 1$. Yet, high magnetization is unlikely to make {\it Guitar} pulsar ($V_\text{NS} \sim 1000$~km/s) sub-fast magnetosonic.

\begin{table}
\centering
\begin{tabular}{c c c c}
 \hline
 $c_s$     & $v_A$       & $M_s$      & $M_A$ \\
 km/s      & km/s        &            &       \\
 \hline
 $\sim 10$ & $\sim 1-10$ & $\sim 100$ & $\sim 100 - 10^3$ \\
 \hline \end{tabular} \caption{Typical ISM parameters: the speed of sound, \Alfven{} velocity, Mach number and \Alfven{} Mach number.} \label{tab:ISM_typical_quantities}
\end{table}

Finally, pulsars and PWNe accelerate particles to high energies. An estimate of the energy and the Lorentz factor being gained in fraction $\eta_\text{acc}$ (acceleration efficiency) of the maximal potential of a pulsar are \citep[see ][]{2002PhRvD..66b3005A}:
\begin{align}
E_{e^\pm,~\text{max}} &=\eta_\text{acc} e \sqrt{\frac{\dot{E}}{c}} = 55~\eta_\text{acc} \dot{E}_{33}^{1/2}~\text{TeV}, \label{eq:max_energy} \\
\gamma_\text{max} &= \frac{E_{e^\pm,~\text{max}}}{m_e c^2} = 1.1 \times 10^8~\eta_\text{acc} \dot{E}_{33}^{1/2}.
\end{align}

Though it reminds of a  Hillas criterion \citep{1984ARA&A..22..425H}, Equation~(\ref{eq:max_energy}) is based on the assumption that electric field of the order of the magnetic field  accelerates particles over some  characteristic scale (magnetic field{} at the light cylinder times the size of the light cylinder). Importantly, the estimate refers  to the potential of the pulsar, {\it not} the PWN. In case of bow-shock PWNe, when the motion remains relativistic both in the unshocked pulsar wind, and post-shock nozzle-like flow in the tail, Equation~(\ref{eq:max_energy}) also applies to the potential of the PWN.

The external  (ISM) magnetic field is a special, fairly subtle issue. First, a reconnection of the external magnetic field  with the PWN fields may allow particles accelerated within the PWN to escape, forming X-ray misaligned filaments (or kinetic jets) \citep{2008A&A...490L...3B,2019MNRAS.485.2041B} -- elongated features protruding away, sometimes ahead,  from the pulsar (in passing,  we note that the prediction that filaments near the Galactic center are powered by pulsars \citep{2019MNRAS.489L..28B} seem to be confirmed by detection of a pulsar in the most prominent node of the ``Snake'' filament  \citep{2024MNRAS.530..254Y}). The pulsar {\it  illuminates} a magnetic field  structure of the ISM by injecting relativistic particles. Streaming motion along magnetic field lines dominates the diffusion, resulting in the perceived collimation.

We stress that kinetic jets are {\it not}  separated hydrodynamic flow, but a collection of high energy particles streaming through mostly undisturbed ISM. In an ambient magnetic field $B_\text{amb}$, particles with energy (\ref{eq:max_energy}) produce synchrotron emission on \citep{1975ctf..book.....L}:
\begin{equation}\label{eq:cutoff_energy}
        E_\text{cutoff} = \alpha \frac{3 \hslash e}{2 c m_{e}} \gamma_\text{max}^{2} B_\text{amb} \sin{\zeta} = 0.50~\text{keV}~B_{\text{amb},-5} \gamma_{\text{max}, 8}^{2} \sin{\zeta},
\end{equation}
Here $\alpha = argmax_\xi \left(\xi \int_\xi^\infty {K_{5/3} (x) dx} \right) \approx 0.29$, $\zeta$ is a pitch angle of the particles in a filament. The latter is largely unknown, despite the recent progress in understanding its' evolution along the filament \citep{2026arXiv260223850S}. In the paper we set $\sin{\zeta} = 1$ for the most conservative estimate on $\eta_\text{acc}$.

By substitution of (\ref{eq:max_energy}) to (\ref{eq:cutoff_energy}), one can estimate the ambient magnetic field:
\begin{equation}\label{eq:B_ism_Xray}
        B_\text{amb} = \frac{2 c^{6} m_{e}^{3}}{3 \alpha \hslash e^{3}} \frac{E_\text{cutoff}}{\eta_\text{acc}^{2} \dot{E}} = 170~\mu \text{G}~\eta_\text{acc}^{-2} \left( \frac{E_\text{cutoff}}{10~\text{keV}}\right) \dot{E}_{33}^{-1}.
\end{equation}

\section{{\it Guitar} is exceptional \label{sec:Guitar_exceptionality}}

After laying out the expectations, we next demonstrate that all observed bow-shock PWN correspond to the general picture, except {\it Guitar}. 

In Table~\ref{tab:filament_compilation} we list the known  kinetic jets. In the last column we give the estimates of the inferred magnetic field  in the ISM, Equation~(\ref{eq:B_ism_Xray}). Except for {\it Guitar}, the inferred magnetic field  is of the order of few $\mu$G. In the case of {\it Guitar} it is nearly hundred times higher, $\sim 100\, \mu$G. Importantly, this high magnetic field is the lower limit, since we assumed the most efficient acceleration, $\eta_{acc} =1$.
\begin{table*}
\centering
\begin{tabular}{l c c c c c c}
 \hline
 Pulsar & $\dot{E}$ & $l_{\text{H}_\alpha} / r_s$ & $l_\text{X-ray} / l_f$ & $E_\text{cutoff}$ & $E_{e^\pm,~\text{max}} (\eta_\text{acc} = 1)$ & $B_\text{min} (\eta_\text{acc} = 1)$ \\
  & $\mathrm{erg\,s^{-1}}$ &  &  & $\mathrm{keV}$ & $\mathrm{PeV}$ & $\mu \mathrm{ G}$ \\
 \hline
 J2225+6535 & $1.2 \times 10^ {33}$ & $100~(800^*)$ & $0.026 \pm 0.006$ & $\gtrsim 10$ & $0.060$ & $144$ \\
 \hline
 J1101-6101 & $1.4 \times 10^ {36}$ &  & $2.1 \pm 0.1$ & $\gtrsim 25$ & $2.0$ & $0.31$ \\
 J2030+4415 & $2.2 \times 10^ {34}$ & $10$ & $0.23 \pm 0.02$ & $\gtrsim 5$ & $0.26$ & $3.9$ \\
 J1509-5850 & $5.1 \times 10^ {35}$ & $30$ & $1.5 \pm 0.3$ & $\gtrsim 7$ & $1.2$ & $0.24$ \\
 J2055+2539 & $5.0 \times 10^ {33}$ &  & $1.21 \pm 0.09$ & $\gtrsim 7$ & $0.12$ & $24$ \\
 J1740+1000 & $2.3 \times 10^ {35}$ &  &  & $\gtrsim 7$ & $0.83$ & $0.53$ \\
 J1957+5033 & $5.3 \times 10^ {33}$ &  &  & $\gtrsim 5$ & $0.13$ & $16$ \\
 \hline
\end{tabular}
\caption{Currently known pulsar filaments \citep{2024ApJ...976....4D}, referred by the pulsar name. The quantities are spin-down power \citep[$\dot{E}$, ANTF catalogue, assuming $I = 10^{45}~\text{g cm}^2$, ][]{2005AJ....129.1993M}, the length of \Ha{} nebula ($l_{\text{H}_\alpha}$, if present) to stand-off distance ($r_s$) \citep{2014ApJ...784..154B}, the length of X-ray PWN ($l_\text{X-ray}$, if present) to the length of the filament ($l_f$) \citep{2024ApJ...976....4D}, the lower limit on the cut-off energy of the filament spectrum ($E_\text{cutoff}$, estimated based on spectra from \cite{2024ApJ...976....4D}; in case of PSR J2225+6535 from \cite{2012ApJ...747...74H}; in case of PSR J1101-6101, based on images and spectra from \cite{2023ApJ...950..177K}), maximum possible energy of the particles ($E_{e^\pm,~\text{max}} (\eta_\text{acc} = 1)$, Equation~(\ref{eq:max_energy})), minimum possible ambient magnetic field strength ($B_\text{min} (\eta_\text{acc} = 1)$, Equation~(\ref{eq:B_ism_Xray})). The latter two quantities are computed assuming perfect acceleration (acceleration efficiency $\eta_\text{acc} = 1$). $^*$ -- stand-off distance and the size of the nebula from \cite{2022ApJ...939...70D}.}
\label{tab:filament_compilation}
\end{table*}

PSR J2225+6535 is a well studied object, with both timing and astrometric data. This, coupled to existing integral field unit observations of the {\it Guitar}, allows us to reconstruct 3D velocity of the pulsar and stand-off distance of the bow shock with unprecedented precision (\ref{sec:reconstruction}) and to make estimates of the possible spin-down power of the pulsar (\ref{subsec:spindown_estimate}). The solutions for $r_s$ and $V_\text{NS}$ do not follow gaussian distribution and we work with them in the form of samples from their distribution (i.e. points in the parameter space; the more points contained in the specified range, the more probable it is). This allows us to propagate uncertainties through the equations. Hereafter the inferred point estimate for the parameter is 50-th percentile in the distribution, uncertainties mark its' differences with 16-th and 84-th percentiles. We denote the range of possible spin-down power with two values, hereafter the quantities inferred with the second one are in brackets. The values are presented in Table~\ref{tab:observational_parameters}.
\begin{table}
\begin{tabular}{c c c c}
 \hline
 $E_\text{cutoff}$, keV & $\dot{E},~10^{33}$~erg/s              & $V_\text{NS}$, km/s & $r_s,~10^{15}$~cm \\
 \hline
 $\gtrsim 10$             & $1.2~(2.3)$    & $780^{+154}_{-100}$ & $1.19^{+0.23}_{-0.16}$ \\
 \hline
\end{tabular}
\caption{Observational parameters of {\it Guitar}  nebula, its filament and PSR J2225+6535. They are cutoff energy of filament X-ray spectrum $E_\text{cutoff}$ (no cutoff in XMM-Newton band \citep{2012ApJ...747...74H}), minimal and maximal estimates for pulsar spin-down power $\dot{E}$ (\ref{subsec:spindown_estimate}), pulsar velocity $V_\text{NS}$, bow shock stand-off distance $r_s$. Note that the last two parameters are strongly correlated (\ref{sec:reconstruction}).}
\label{tab:observational_parameters}
\end{table}

\begin{figure*}
        \centering
        \includegraphics[width=\textwidth]{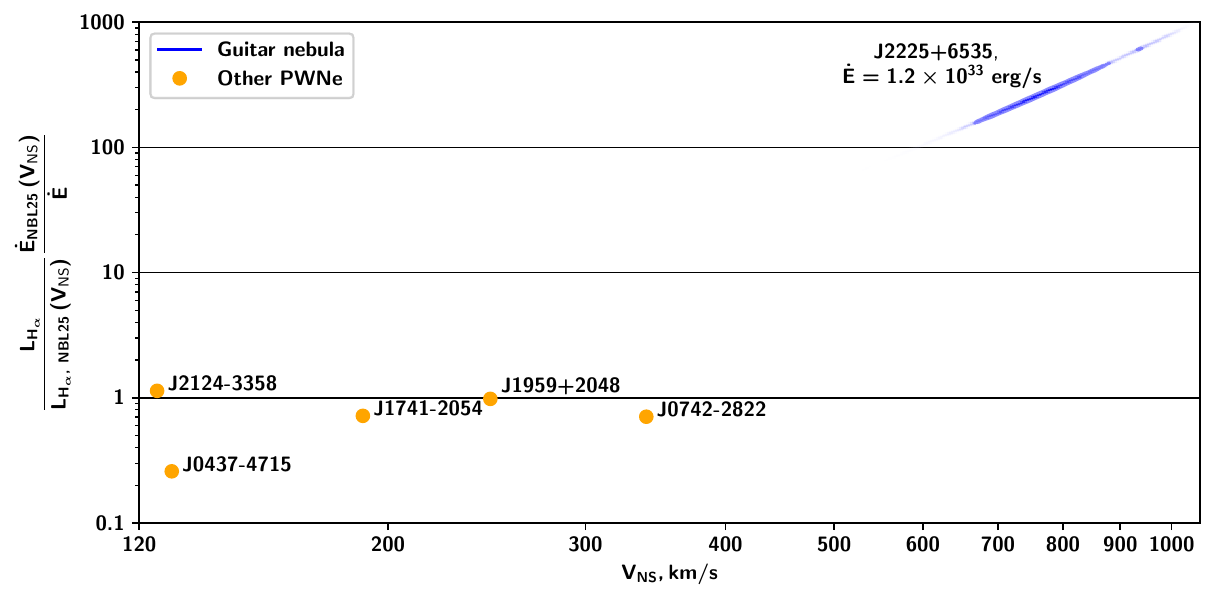}
        \caption{\Ha{} luminosity of PWNe normalized to NBL25 model. For {\it Guitar} we show the density of samples from the distribution of parameters. The contour corresponds to 0.393 confidence level ($1 \sigma$ in case of 2D Gaussian). PWNe are referred to by the pulsar name}
        \label{fig:PWNe_luminosity}
\end{figure*}

{\it Guitar} turns out to be exceptional in \Ha{} luminosity. In Figure~\ref{fig:PWNe_luminosity} we present the comparison of the PWNe luminosity to NBL25 models. The analysis is similar to NBL25, but here we added {\it Guitar} as a set of samples and normalized the luminosity to the one in models. As a result, {\it Guitar} is $\sim 300$ times brighter than expected. It points to the unique properties of Guitar, because assumptions of NBL25, namely weak ambient magnetic field and suitability of ideal hydrodynamics may not apply to it. This  can be a consequence of high ambient density without an expected reduction in the nebula size.

Even with assumption of an extreme intrinsic particle acceleration (central pulsar accelerates particles to maximal available potential), which is remarkable by itself, the ambient medium must have exceptionally high magnetic field, making pulsar's local ISM unique. This conclusion is amplified by the exceptional brightness of \Ha{} nebula. Section~\ref{sec:Bism_bowshock_config} focuses on the modelling of the apex of the Guitar nebula, disentangling properties of extreme particle acceleration and  unique parameters of the medium (summarized in Table{\ref{tab:model_parameters_bootstrap}}).

\section{Self-consistent model of BSPWN in external magnetic field. \label{sec:Bism_bowshock_config}}

To assess the physical parameters at the Guitar nebula apex, we construct an analytical model accounting for observed morphology, as well as filament visibility at X-ray energies, extending Section~\ref{sec:BSPWNe_description}. It incorporates the effect of the ambient magnetic field on the pressure of the medium and the synchrotron emission. As the result, the model is self-consistent and algebraic.

\subsection{General equations and assumptions\label{subsec:bowshock_config_general_equations}}

Consider the dynamics of interaction between interstellar gas and pulsar wind. Normally, the gas passing the bow shock is fully ionized before reaching contact discontinuity, so its momentum is fully deposited to the flow exerting ram pressure $\mu_\text{ISM} n_\text{ISM} m_p V_\text{NS}^2$ at the apex (we expressed the ISM density through the number density ($n_\text{ISM}$) and the mean molecular mass of ISM gas ($\mu_\text{ISM}$)). In quasistationary configuration, at the contact discontinuity it is in equilibrium with pulsar wind pressure:
\begin{equation}\label{eq:pressure_wind}
        p_w = \frac{\dot{E}}{4 \pi c r_{s}^{2}}.
\end{equation}
This results in Equation~(\ref{eq:Rs}). 

For {\it Guitar}, the following things are different:
\begin{itemize}
        \item In {\it Guitar} nebula only a small part of ISM gas is being ionized between the bow shock and the contact discontinuity, as shown in Section~\ref{subsec:ionization_structure} and  \ref{sec:Ionization}. Only a small part of the momentum of ISM is deposited, since pulsar wind interacts with ionized fraction. Since the flow is mostly collisionless, we assume deceleration of the ions before contact discontinuity is negligible, and assign them velocity $V_\text{NS}$. We denote their number density as $n_{i,c}$. The ram pressure on the contact discontinuity becomes: 
        \begin{equation}\label{eq:pressure_ram}
                p_\text{ram} = \mu_\text{ISM} n_{i,c} m_p V_\text{NS}^2.
        \end{equation}
        By the structure of the expression, if our assumption does not hold, $n_{i,c}$ still denotes the number density of atoms in ISM that are ionized before reaching contact discontinuity;

        \item The magnetic field in the ambient medium is extremely high (see Table~\ref{tab:filament_compilation}). We assess its influence on the dynamics via magnetic pressure 
        \begin{equation}\label{eq:pressure_magnetic}
                p_B = \frac{B_\text{amb}^2}{8 \pi}
        \end{equation}
        (for $B_\text{amb}$ and hence filament itself being perpendicular to pulsar velocity, which is the case for {\it Guitar}).
\end{itemize}

In this work we fix $\mu_\text{ISM} = 1.26$, one can account for another value by renormalizing $n_{i,c}$. By treating all quantities in a self-consistent manner, we infer connections between $\dot{E}$, $E_\text{cutoff}$, $r_s$ and $V_\text{NS}$. Firstly, similarly to Equation~(\ref{eq:Rs}), we get:
\begin{equation}\label{eq:B_ism_pressure}
        B_\text{amb} = \sqrt{2 \frac{\dot{E}}{c r_{s}^{2}} - 8 \pi \mu_\text{ISM} n_{i,c} m_p V_\text{NS}^2}.
\end{equation}

In the physical picture outlined in Section~\ref{sec:BSPWNe_description}, escaping particles highlight an ambient magnetic field. Thus, magnetic field compressing the bow shock (Equation~(\ref{eq:B_ism_pressure})) is the same field in which the emission occurs (Equation~(\ref{eq:B_ism_Xray})). By combining corresponding equations and eliminating the magnetic field from them, we get the relation between the spin-down power and the cutoff energy of the kinetic jet spectrum:
\begin{equation}\label{eq:Espindown_Emax_dimensional}
       \sqrt{2 \frac{\dot{E}}{c r_{s}^{2}} - 8 \pi \mu_\text{ISM} n_{i,c} m_p V_\text{NS}^2} = \frac{2 E_\text{cutoff} c^{6} m_{e}^{3}}{3 \alpha \dot{E} \eta_\text{acc}^{2} \hslash e^{3}}.
\end{equation}

\subsection{Renormalization and solution}

We introduce new dimensionless variables $(X,~A)$:
\begin{align} 
        \frac{p_w}{p_{ram}} &= X \ge 1, \label{eq:dimensionless_variables_X} \\
        \frac{p_{w}}{p_{B}} &= \frac{27 X^{3}}{2 A^{2}} \ge 1. \label{eq:dimensionless_variables_A}
\end{align} 
$X$ measures the ratio between pulsar wind pressure and ram pressure of the upstream ISM, and thus its value over 1 shows an excess of pressure, which magnetic pressure has to cover. If in Equation~(\ref{eq:dimensionless_variables_A}) we fix $p_w$ and $p_{ram}$, we infer $p_B \propto A^2$ and thus $B \propto A$. The meaning of $A$ is a measure of an ambient magnetic field strength.

To analyze and solve equation~(\ref{eq:Espindown_Emax_dimensional}) we convert $\left(\dot{E},~\frac{E_\text{cutoff}}{\eta_\text{acc}^{2}}\right)$ to the units $\left( k_X, k_A \right)$, constructed in a way that $(X,~A)$ are their numerical values: 
\begin{align}
        \dot{E}   &= k_X X, \label{eq:substitution_spindown} \\
        \frac{E_\text{cutoff}}{\eta_\text{acc}^{2}} &= k_A A. \label{eq:substitution_exmax}
\end{align}
The expressions for the introduced constants are:
\begin{align}
        k_X &= 4 \pi c m_p \mu_\text{ISM} n_{i,c} V_{NS}^{2} r_{s}^{2} = \notag \\
            &= 0.79 \times 10^{33}~\text{erg/s}~\left(\frac{\mu_{c}}{1.26}\right) n_{c, -1} V_{\text{NS}, 8}^2 r_{s, 15}^2, \label{eq:k_X} \\
        k_A &= \frac{8 \sqrt{3} \pi^{\frac{3}{2}}}{3} \frac{\alpha \hslash e^{3} m_p^{\frac{3}{2}}}{m_{e}^{3} c^{5}} \mu_\text{ISM}^{\frac{3}{2}} n_{i,c}^{\frac{3}{2}} V_{NS}^{3} r_{s}^{2} = \notag \\
            &= 2.9~\text{keV}~\left(\frac{\mu_{c}}{1.26}\right)^\frac{3}{2} n_{c, -1}^\frac{3}{2} V_{\text{NS}, 8}^3 r_{s, 15}^2. \label{eq:k_A}
\end{align}

\begin{figure}
	\centering 
        \includegraphics[width=0.5\textwidth]{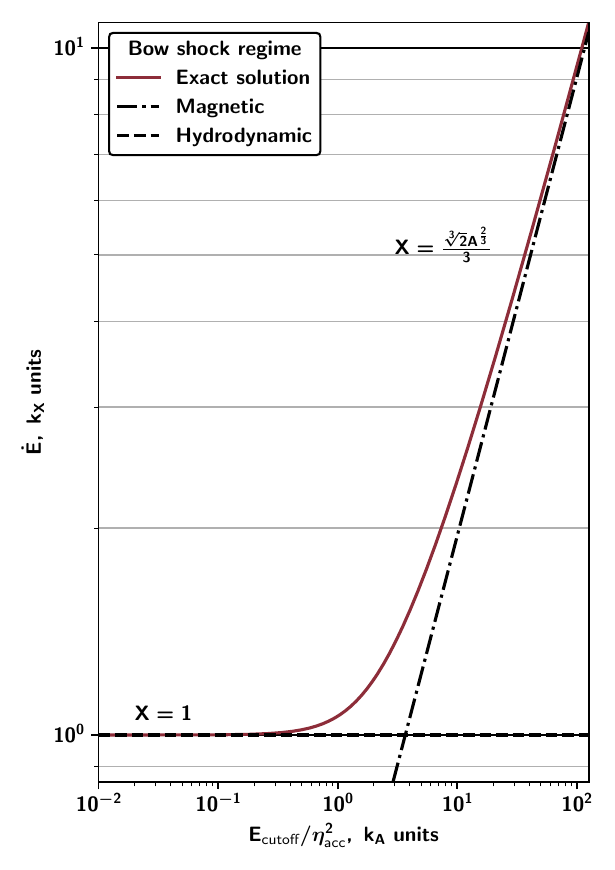}
        \caption{Dependency between dimensionless cutoff energy of kinetic jet spectrum ($A$) and spin-down power ($X$) of the pulsar. The plot represents different regimes of bow shock and follows the solution of Equation~(\ref{eq:Espindown_Emax_dimensionless})}        
        \label{fig:X_A_plot}
\end{figure}

Equation~(\ref{eq:Espindown_Emax_dimensional}) in dimensionless form becomes:
\begin{equation}\label{eq:Espindown_Emax_dimensionless}
        \sqrt{X - 1} = \frac{\sqrt{6} A}{9 X}.
\end{equation}
We infer the analytical solution of equation~(\ref{eq:Espindown_Emax_dimensionless}): 
\begin{align}
        X &= \frac{1}{3} \left(K^\frac{1}{3} + 1 + K^{-\frac{1}{3}} \right), \label{eq:X_K_solution} \\
        \text{where } K &= A^{2} + A \sqrt{A^{2} + 2} + 1. \label{eq:D_A_solution} 
\end{align}
It is presented in fig.~\ref{fig:X_A_plot} with its asymptotes.

If magnetic field in ISM is negligible, the filament cannot be seen at X-ray energies, $A = 0$. In this case, 
\begin{equation}\label{eq:hd_regime}
X = 1.
\end{equation}
We call this hydrodynamic regime, since $p_{ram} = p_w$ -- ram pressure fully counteracts pulsar wind pressure and no magnetic pressure is present.

If exponential cutoff in kinetic jet spectrum is observed at very high energies, $A \gg 1$. In this case, 
\begin{equation}
        X = \frac{\sqrt[3]{2}}{3} A^\frac{2}{3}. \label{eq:X_highK_solution}
\end{equation} 
We call this magnetic regime, since $p_B = p_w$. Ram pressure of ISM in this case is negligible in the direction perpendicular to the magnetic field. However, in real nebulae the pressure of the wind in the regions where the normal to the shock is parallel to the ISM magnetic field direction must be in dynamic equilibrium with the ambient ram pressure. The solution for the bow shock as a whole in a highly magnetized medium must consider this effect. This can be done in the framework of computational relativistic magnetohydrodynamics \citep{2019MNRAS.484.4760B, 2019MNRAS.485.2041B} and is out of scope of this work. Equation~(\ref{eq:X_highK_solution}) can be used as a limit.

\subsection{Implications for {\it Guitar}}

To analyze the model we build a parametric solution for the spin-down power from ISM number density given the cutoff energy of filament X-ray spectrum (directly with Equations~\ref{eq:substitution_spindown} -- \ref{eq:k_A} and \ref{eq:X_K_solution} -- \ref{eq:D_A_solution}):
\begin{equation}\label{eq:parametric_expression}
        \dot{E} = f\left(n_{i,c}, \frac{E_\text{cutoff}}{\eta_\text{acc}^{2}} \right).
\end{equation} 
For that, we build the dependence of the solution for the spin-down power on $n_{i, c}$ (number density of ions on the contact discontinuity, see Section \ref{subsec:bowshock_config_general_equations} for the discussion), which appears in Equations~\ref{eq:k_X} and~\ref{eq:k_A}. $\dfrac{E_\text{cutoff}}{\eta_\text{acc}^{2}}$ transitions into $A$, which defines $X$ and in turn $\dot{E}$.

\begin{figure}
	\centering 
        \includegraphics[width=0.5\textwidth]{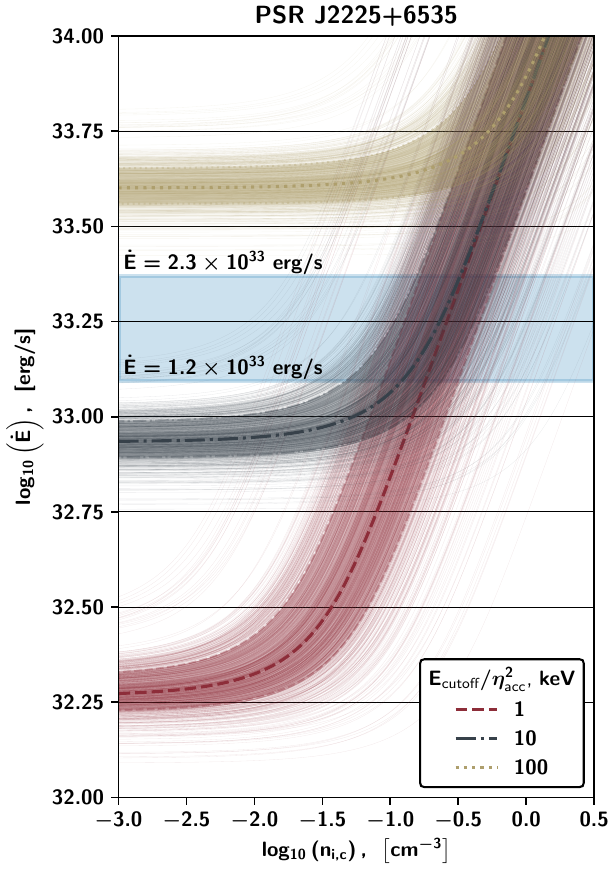}
        \caption{Dependence between ion number density on the contact discontinuity ($n_{i,c}$) and spin-down power ($\dot{E}$) for Guitar PWN. We plot 16-th, 50-th and 84-th percentiles by dashed, dash-dotted and dotted lines for the values 1, 10 and 100~keV of the filament spectrum cutoff energy divided by acceleration efficiency squared. For each value we fill the area between 16-th and 84-th percentile with color and plot 1000 random samples from the distribution of the solutions. Here $\mu_\text{ISM} = 1.26$, affects only $n_{i,c}$ normalization. Horizontal lines around the shaded area are the minimal ($\dot{E}_\text{min}$) and the maximal ($\dot{E}_\text{max}$) estimates for the spin-down power of PSR J2225+6535 (see Section~\ref{subsec:spindown_estimate}).}
        \label{fig:dotE_spindown_rho_ism}
\end{figure}

We present the result in Figure~\ref{fig:dotE_spindown_rho_ism} given various preassigned values of cut-off energy. Contrary to Figure~\ref{fig:X_A_plot}, on the ``plateau'' (lower left part of the curves) magnetic pressure is dominant and the required spin-down power is little affected by the number density of the ions (magnetic regime). In the same time, in the upper right part of the plot the curves form a power law. There ram pressure is dominant and Equation~(\ref{eq:Rs}) holds (hydrodynamic regime). 

\subsubsection{Minimum cut-off energy of pulsar wind particle spectrum and maximum ambient magnetic field strength}

The most conservative estimate on the acceleration efficiency is for the magnetic regime, where the nebula has the smallest restriction on the spin-down power (see Figure~\ref{fig:dotE_spindown_rho_ism}). In this case,
\begin{equation}\label{eq:efficiency_estimate}
        \eta_\text{acc} = \frac{\sqrt[4]{2} \sqrt{3}}{3} \frac{ c^{\frac{13}{4}} m_{e}^{\frac{3}{2}} }{\sqrt{\alpha \hslash} e^{\frac{3}{2}}} \frac{\sqrt{r_{s} E_\text{cutoff}}}{\dot{E}^{\frac{3}{4}}} \gtrsim 0.76^{+0.07}_{-0.05}~\left( 0.47^{+0.04}_{-0.03} \right)
\end{equation}
This is much larger than previously expected value of $\approx 0.1$. With this, following Equation~(\ref{eq:max_energy}), we estimate maximum particle energy in the filament to be $E_{e^\pm,~\text{max}} \gtrsim 46^{+4}_{-3}~\left( 40^{+4}_{-3} \right)$~TeV. It is close to $\approx 61~\left( 84 \right)$~TeV, which would be the case for ideal acceleration, i.e. unity efficiency.

The lowest possible value of the acceleration efficiency requires the largest possible value for the magnetic field strength. Namely, for the magnetic regime,
\begin{equation}\label{eq:B_estimate_magnetic}
        B_\text{amb} \le \frac{1}{r_{s}} \sqrt{\frac{2 \dot{E}}{c}} \approx 241^{+37}_{-39}~\left( 332^{+51}_{-54} \right)~\mu \text{G}.
\end{equation}
For a weaker ISM magnetic field, the acceleration efficiency must be even higher.

\subsubsection{Maximum number density of the nuclei on the contact discontinuity and minimum ambient magnetic field strength}

As is seen from Figure~\ref{fig:dotE_spindown_rho_ism}, the requirement on the available pulsar spin-down budget grows with $n_{i,c}$. From Equation~(\ref{eq:Espindown_Emax_dimensional}), we have: 
\begin{align}\label{eq:mhd_regime_n_ism}
n_{i,c} &= \frac{\dot{E}}{4 \pi V_\text{NS}^{2} \mu_\text{ISM} c m_{p} r_{s}^{2}} - \\ \notag 
             &- \frac{E_\text{cutoff}^{2} c^{12} m_{e}^{6}}{18 \pi V_\text{NS}^{2} \alpha^{2} \dot{E}^{2} \eta_\text{acc}^{4} \hslash^{2} \mu_\text{ISM} e^{6} m_{p}} = \\ \notag
             &= V_{\text{NS}, 8}^{-2} \left( \frac{\mu_\text{ISM}}{1.26} \right)^{-1} \times \\ \notag 
             &\times \left( 0.126~\dot{E}_{33} r_{s, 15}^{-2} - 0.056~\eta_\text{acc}^{-4} \left( \frac{E_\text{cutoff}}{10~\text{keV}}\right)^{2} \dot{E}_{33}^{-2}\right)  ~\text{cm}^{-3},
\end{align}
For Guitar nebula we infer $n_{i,c} \approx 0.12^{+0.12}_{-0.07} \left( \frac{\mu_\text{ISM}}{1.26} \right)^{-1}~\text{cm}^{-3}$  $\left( 0.33^{+0.25}_{-0.17} \left( \frac{\mu_\text{ISM}}{1.26} \right)^{-1}~\text{cm}^{-3}\right)$ for $\eta_\text{acc} = 1$. For minimal possible acceleration efficiency of 0.76, $n_{i,c} \rightarrow 0$.

In this scenario, we can also estimate the lower limit for ISM magnetic field strength. From Equation~(\ref{eq:B_ism_Xray}), for $\eta_\text{acc} = 1$, we have
\begin{equation}\label{eq:mhd_regime_B_ism}
        B_\text{amb} = \frac{2 c^{6} m_{e}^{3}}{3 \alpha \hslash e^{3}} \frac{E_\text{cutoff}}{\dot{E}} \gtrsim 140~(74)~\mu \text{G}.  
\end{equation}

\subsection{Summary on the inferred parameters}
Figure~\ref{fig:dotE_spindown_rho_ism} shows that the requirement on the spin-down power is quite tight. The particle accelerator has to be near perfect. The presence of a cutoff at energies about 10 -- 20~keV might be checked with the NuSTAR telescope \citep[see, for example,][]{2023ApJ...950..177K}. The detection of a harder cutoff would mean that PSR J2225+6535 would have to be a more powerful pulsar and it could put restrictions on the NS EOS. Altogether, the parameters are summarized in Table~\ref{tab:model_parameters_bootstrap}.
\begin{table}
\begin{tabular}{l c}
 \hline
 $\eta_\text{acc}$                & $\gtrsim 0.76^{+0.07}_{-0.05}~\left( 0.47^{+0.04}_{-0.03} \right)$  \\
 $E_{e^\pm,~\text{max}}$,~[TeV]   & $46^{+4}_{-3}~\left( 40^{+4}_{-3} \right) \lesssim E_{e^\pm,~\text{max}} \lesssim 61~\left( 84 \right)$ \\
 $B_\text{amb}$,~[$\mu$G]         & $140~\left( 74 \right) \lesssim B_\text{amb} \lesssim 241^{+37}_{-39}~\left( 332^{+51}_{-54} \right)$ \\
 $n_{i,c},~[\text{cm}^{-3}]$ & $\lesssim 0.12^{+0.12}_{-0.07}~\left( 0.33^{+0.25}_{-0.17} \right)$  \\
 \hline
\end{tabular}
\caption{Parameters of the model: the efficiency of particle acceleration in PSR J2225+6535 wind ($\eta_\text{acc}$), maximum energy of the wind particles ($E_{e^\pm,~\text{max}}$), ambient magnetic field strength ($B_\text{amb}$) and number density of ions at contact discontinuity ($n_{i,c}$). The last value is calculated assuming $\mu_\text{ISM} = 1.26$. Values outside brackets and inside brackets correspond to $\dot{E}_\text{min}$ and to $\dot{E}_\text{max}$ respectively (\ref{subsec:spindown_estimate}).}
\label{tab:model_parameters_bootstrap}
\end{table}

\section{The scenario \label{sec:scenario}}

In Sections~\ref{sec:Guitar_exceptionality} and \ref{sec:Bism_bowshock_config} we showed that properties of {\it Guitar} nebula are exceptional --  they cannot be explained by the conventional assumptions about the properties of the local ISM (see \ref{sec:mag_self_amp}). Next, we suggest an explanation: PSR J2225+6535 crosses  a long-predicted region of an old SNR with  exceptionally high magnetic field.

\subsection{Pulsar inside the unrelated SNR}

\begin{figure}
	\centering 
        \includegraphics[width=0.5\textwidth]{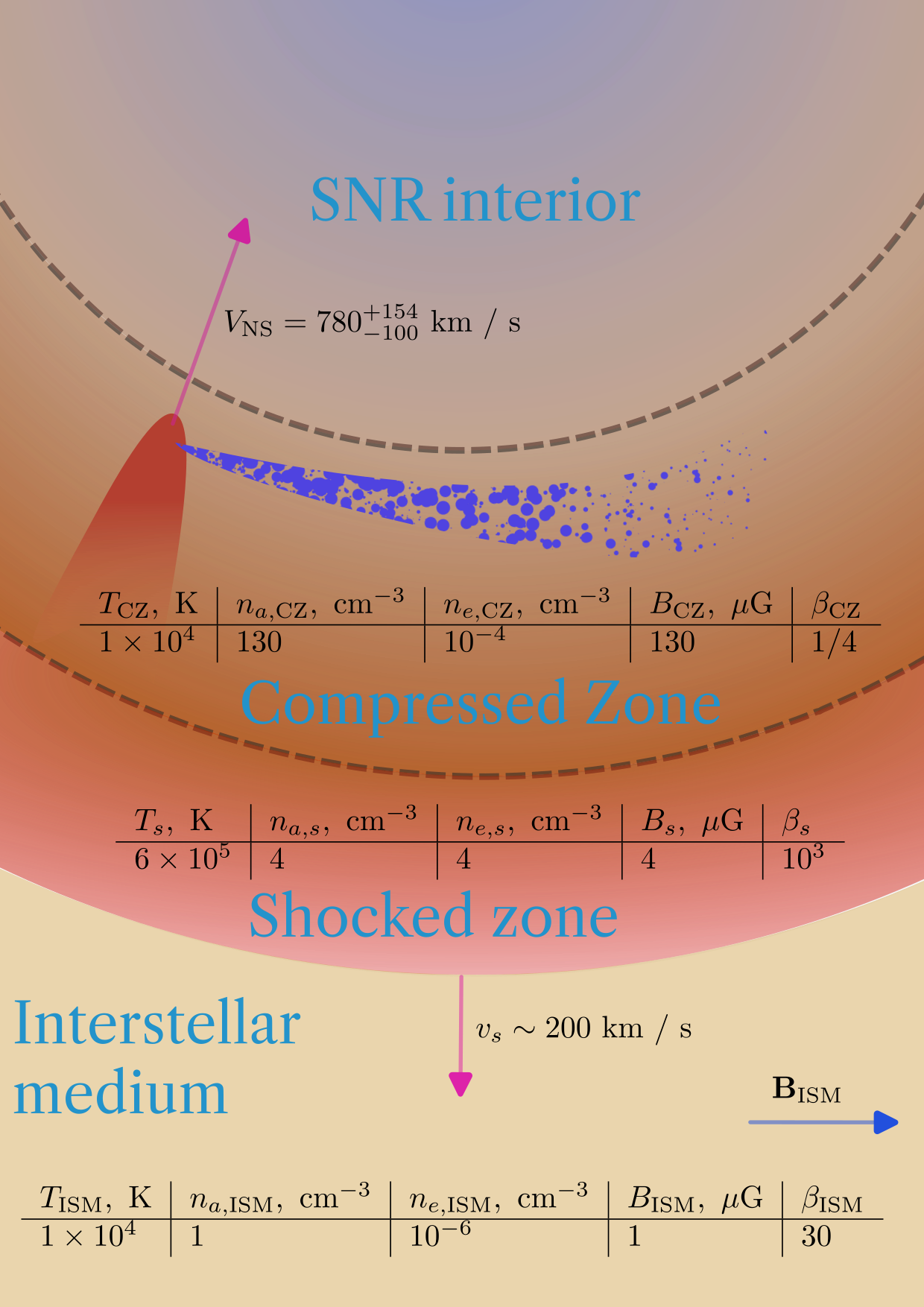}
        \caption{Cartoon of the proposed scenario: a pulsar moves through an old SNR shell. The parameters are temperature ($T$), number density of nuclei ($n_a$), number density of electrons ($n_e$), magnetic field strength ($B$), plasma beta parameter ($\beta$)}
        \label{fig:cartoon}
\end{figure}

The general picture is introduced in Figure~\ref{fig:cartoon}. We suggest that the pulsar is now moving through an old SNR, which is currently in the pressure-driven snowplow regime \citep{1972ApJ...178..159C,1974ApJ...188..501C}. At the current SNR age, approximately 30-50 kyrs, the hot tenuous central part sustains relatively high pressure, approximately at the Sedov level, while the denser shell is in the cooling snowplow regime. This regime is usually referred to as ``pressure-driven snowplow'' \citep{Chevalier77}.

In particular, 
\begin{itemize}
\item A supernova unrelated to PSR J2225+6535 exploded $\sim 30-50$ kyrs ago, in a common warm neutral  medium ($n_{\text{ISM}} \sim 1~\text{cm}^{-3}$, $B_\text{ISM} \sim 1  \mu$G, $\beta_\text{ISM} \sim 30$, ionization degree $\chi_\text{ISM} \sim 10^{-6}$). SNR expanded to $R_\text{front} \sim 20-50$ pc;
\item At the SNR forward shock (hereafter denoted by index s), the medium is ionized, magnetic field is compressed ($\beta_s \sim 10^3$, $\chi_s \sim 1$);
\item The shocked shell is in a strongly cooling regime, losing thermal pressure to radiation. At the same time, a hot and rarified central part keeps high pressure. As a result, shell material is compressed, together with the frozen-in magnetic field;
\item The cooled "compressed zone" (CZ) becomes cold, mostly neutral and sandwiched between hot interior and hot shocked exterior zones. It is the pressure of the  magnetic field  that stabilizes the compression. The thickness of the shell is $(1 - 2) \times10^{17}$~cm;
\item PSR J2225+6535 happens to traverse now the high density, high magnetic field CZ.
\end{itemize}

\subsection{Formation of the SNR shell compressed zone -- ambient medium for the PWN}

The old SNR should be quite large $R_\text{front} \gtrsim 25$~pc. The snowplow stage is determined by effective radiative cooling at the shock, which takes place at $6\times 10^5 \mbox{ K}\ge T_s \ge 10^4$~K \citep[see for review][]{Chevalier77}. The temperature at the shock can be estimated as 
\begin{equation}
    T_s=\frac{3 \mu_s m_p v_s^2}{16 k_B},
    \label{eq:Tx}
\end{equation}
here $k_B$, $m_p$ are the Boltzmann constant and proton mass, respectively, and $\mu_s=0.61$ is the mean molecular weight. The velocity of shock that triggers effective radiative cooling is $v_s \approx 200$~km/s (it can be different due to partial proton-electron equilibration, the more sophisticated expression is Equation~(\ref{eq:Te_noneq})).

The interaction of the old SN with ISM compresses density and magnetic field by four times. Here for clarity we introduce the new variable -- the number density of nuclei ($n_a$). For initial values of $n_{a,\text{ISM}} = 1~\text{cm}^{-3}$ and $B_\text{ISM} = 1~\mu$G (perpendicular to the shock normal), after the shock $n_{a,s} = 4~\text{cm}^{-3}$ and $B_s = 4 \mu$G. The temperature is $T_s = 6 \times 10^{5}$~K, and corresponding plasma pressure $P_{s,T} = 7 \times 10^{-10}~n_{a,\text{ISM},0}$~Ba. The initial plasma-beta at the shock is $\beta_s = 10^3~n_{\text{ISM},0} B_{\text{ISM},-6}^{-2}$.
 
In the snowplow regime, the shocked matter has enough time to cool down to $T_\text{CZ} = 10^4$~K. Inside the SNR, from the shock to the interior, the plasma is in dynamical balance, so  the total pressure can be treated as constant. This corresponds to a maximum compression ratio of 480 times relative to ISM density (4 (shock) $\times 60$ (temperature) $\times 2$ (ionization)). The pressure of the compressed magnetic field $P_{CZ, m}$ will limit compression on $P_{CZ, m} = P_{s,T}$. The maximal magnetic field in the CZ can then be estimated as
\begin{equation}
    B_\text{CZ,~max} = \sqrt{8 \pi (n_{a,s} + n_{e,s}) k_B T_s} = 130~n_{\text{ISM},0}^{1/2}~\mu \text{G}.
        \label{eq:Bsmax}
\end{equation}
Since the magnetic flux is conserved, the gas is compressed to:
\begin{equation}
    n_\text{CZ,~max} = \frac{B_\text{CZ,~max}}{B_s} n_s = 130~n_{\text{ISM},0}^{3/2}~\text{cm}^{-3}.
        \label{eq:nsmax}
\end{equation}
   
Curiously plasma $\beta_\text{CZ} \sim 1/4$ can be relatively small, but it is supported by the external pressure of the hot plasma on both sides. The gas cools and recombines forming a low ionization ($\chi  = n_{e,\text{CZ}} / n_\text{CZ} \approx 10^{-6}$) shell.

Radiatively cooling SNR shell is subjected to various instabilities (like Rayleigh--Taylor (RT), or Richtmyer--Meshkov (RM)), which are sensitive to bending perturbations \citep{2016MNRAS.459.2188B}. They can be suppressed either if neutral gas comprises a large part of the shell or, more realistically, in presence of external magnetic field with a strength of several $\mu$G perpendicular to the shock normal, since magnetic pressure counteracts bending \citep{2021AstL...47..746S}.

\subsection{Ionization after the bow shock \label{subsec:ionization_structure}}
    
The pulsar of {\it Guitar} nebula enters the CZ, its' wind interacts with the ionized component and forms a bow shock. The electron number density inside the bow shock has the following structure (the detailed treatment is considered in \ref{sec:Ionization}):
\begin{equation}
    n_e = \chi_\text{CZ} n_\text{CZ,~max} \exp\left( 0.3 n_\text{CZ,~max} C_{e} \frac{r_s}{V_\text{NS}} \right),
    \label{eq:nes}
\end{equation}
where $C_{e}\sim 10^{-8} \mbox{ cm}^3\mbox{ s}^{-1}$ is ionization coefficient, and $\approx 0.3 r_s$ is the distance between the bow shock and the contact discontinuity \citep{2019MNRAS.484.4760B}. The electron number density exponentially grows up to $n_e\approx 0.1~\text{cm}^{-3}$, which is in agreement with the ion number density at the contact discontinuity (Table~\ref{tab:model_parameters_bootstrap}). The gas in the tail is still being ionized and increases the emissivity of the tail many times (about 300 times compared to the NBL25 hydrodynamic model prediction, Figure~\ref{fig:PWNe_luminosity}).

\subsection{Dispersion measure \label{subsec:DM}}

PSR J2225+6535 has a dispersion measure ($DM$) of $36.4~\text{cm}^{-3}\text{pc}$ \citep{2005AJ....129.1993M}, which corresponds to a distance of about 1.86~kpc \citep{2002astro.ph..7156C,2017ApJ...835...29Y}. The parallax distance is $D = 0.83^{+0.16}_{-0.11}$~kpc (\ref{sec:reconstruction}), to which $DM=13.6^{+2.7}_{-1.9}~\text{cm}^{-3}\text{pc}$ corresponds, and the rest of $\Delta DM\approx 23 ~\text{cm}^{-3}\text{pc}$ can be associated with the environment around the nebula itself.

In our scenario the part of dispersion measure may be attributed to the fully ionized part of the supernova shell, which is upstream of the cooling wave, it has a thickness of: 
\begin{equation}\label{eq:front_thickness}
H_s = \frac{\kappa}{3} \frac{\rho_\text{ISM}}{\rho_{s}} R_\text{front}\approx 1\left(\frac{\kappa}{1/2}\right) \left(\frac{R_\text{front}}{25~\text{pc}}\right) \;\rm pc,
\end{equation}
where $\kappa$ is the mass fraction of the shocked zone to all ISM gas swept by the SNR; $\rho_\text{ISM}$ and $\rho_{s}$ are gas density upstream and downstream of the shockwave, $\rho_{s} / \rho_\text{ISM} = 4$; $R_\text{front}$ is the SNR radius. If the gas consists mostly of hydrogen, downstream electron number density is $n_{e,s} = \rho_{s} / m_p$ and the dispersion measure normal to the SNR shell is:
\begin{equation}
        DM_{H_s} = H_s n_{e,s} =  \frac{\kappa}{3} n_\text{ISM} R_\text{front} \approx 4~ \left(\frac{\kappa}{1/2}\right) n_{\text{ISM}, 0} \left(\frac{R_\text{front}}{25~\text{pc}}\right)~\text{cm}^{-3}\text{pc},
                        \label{eq:DMf}
\end{equation}
where $n_\text{ISM}$ is the upstream hydrogen number density, which we assume to be the typical density of the warm component of Galactic ISM.

The line of sight is in ionized zone for $l_s \le l_{s,\text{max}} \approx \sqrt{2 H_s R_\text{front}}$. Thus, DM can be boosted by geometric effect up to:
\begin{align}
DM_{l_{s,\text{max}}} &= DM_{H_s} \frac{l_{s,\text{max}}}{H_s} \approx \sqrt{\frac{24}{\kappa}} DM_{H_s} = \notag \\
                      &= 28  \left(\frac{\kappa}{1/2}\right)^\frac{1}{2} n_{\text{ISM}, 0} \left(\frac{R_\text{front}}{25~\text{pc}}\right)~\text{cm}^{-3}\text{pc}. \label{eq:DMmax}
\end{align}
We should note that the observed $\Delta DM < DM_{l_{s,\text{max}}}$ and is in agreement with the proposed scenario.

\subsection{Rotation measure \label{ssec:RM}}

The rotation measure (RM) of the pulsar can be significantly affected by the SNR shell \citep{2022MNRAS.515.4217B}. It can be calculated as:
\begin{equation}
        RM = \frac{e^3}{2 \pi \left( m_e c^2 \right)^2} \int_0^D n_e B \cos \psi dR.
\end{equation}

In the considered scenario we infer the part of RM inside the SNR shell:
\begin{align}
        \Delta RM &= \frac{e^3}{2 \pi \left( m_e c^2 \right)^2} \cos \psi \left[ n_{e,s} B_s l_s + n_{e,\text{CZ}} B_\text{CZ} l_\text{CZ} \right] = \notag
        \\
        &= 81~\cos \psi \left( \frac{\Delta DM}{25~\text{cm}^{-3} \text{pc}} \right) \left( \frac{B_s}{4~\mu \text{G}}\right) (1 + x)~\text{rad}~\text{m}^{-2}, \\ \notag
        \text{where}~x &\equiv \frac{n_{e,\text{CZ}}}{n_{e, s}} \frac{B_\text{CZ}}{B_s} \frac{l_\text{CZ}}{l_s}.
\end{align}
Here $\psi$ is the angle between photon travel direction and the magnetic field, i.e. between the filament and the line of sight.

For PSR J2225+6535, $RM = -14.7 \pm 0.7~\text{rad}~\text{m}^{-2}$ \citep{2015MNRAS.453.4485F}. Let us assume $RM = \Delta RM$. In order not to put extreme constraints on $\cos \psi$, $x \lesssim 1$ is required. $\frac{B_\text{CZ}}{B_s} = \frac{\rho_\text{CZ}}{\rho_s}$ and $\frac{n_{e,\text{CZ}}}{n_{e,s}} \approx \chi_\text{CZ} \frac{\rho_\text{CZ}}{\rho_s}$. To allow the path of the light to vary through the full possible range $0 \le \frac{l_\text{CZ}}{l_s} \le \sqrt{\frac{\rho_s}{\rho_\text{CZ}}}$, we thus constrain the ionization degree in the CZ to be $\chi_\text{CZ} < \left( \frac{\rho_s}{\rho_\text{CZ}} \right)^{\frac{3}{2}} \approx 8 \times 10^{-3}$ (for $\frac{\rho_s}{\rho_\text{CZ}} \approx \frac{1}{25}$). This corresponds to $n_{e,\text{CZ}}< 0.8\text{ cm}^{-3}$.

\section{Discussion and Conclusions\label{sec:conclusions}}

{\it Guitar} nebula and PSR J2225+6535 have  exceptionally well measured parameters - this allows detailed quantitative measurements both  of  the particle acceleration  physics within the nebula, and  properties of the external medium.   In this work we  construct a comprehensive model that includes both the high energy emission of the kinetic jets, as well as bright \Ha{}  emission. Our fitted parameter for the 
 ambient magnetic field ($B_\text{amb} > 140 ~(74)~\mu$G)
 and \Ha{} luminosity $ \left( \frac{L_{\text{H}_\alpha}}{L_{{\text{H}_\alpha},~\text{NBL25}}} \frac{\dot{E}_\text{NBL25}}{\dot{E}} \sim 300\right)$ are much higher, than expected. We connected \Ha{} morphology and properties of X-ray kinetic jet in self-consistent analytical model, constraining the physical parameters of pulsar wind particles and the gas on the contact discontinuity (Table~\ref{tab:model_parameters_bootstrap}). It occurs PSR J2225+6535 accelerates pulsar wind particles up to the maximal potential drop (acceleration efficiency $\eta_\text{acc} \gtrsim 0.76^{+0.07}_{-0.05}~\left( 0.47^{+0.04}_{-0.03} \right)$), which makes it an {\it extreme accelerator}.

We propose the scenario incorporating the extreme properties of {\it Guitar} nebula (Section~\ref{sec:scenario}). PSR J2225+6535 passes through an unusual region of high magnetic field, $B_\text{CZ, max} = 130~\mu$G, produced by an old unrelated SNR. This compressed zone ($n_\text{CZ, max} = 130~\text{cm}^{-3}$) consists of the cooled ($\sim 10^4$~K) and highly magnetized ($\beta_\text{CZ} \sim 1/4$) gas, supported by hot SNR zones of a forward shock and an interior. Only a small portion of the gas is being ionized after passing the bow shock of {\it Guitar} nebula, making the flow in the PWN mostly collisionless. As we model the electron number density, it grows exponentially from the value corresponding to the mostly neutral CZ (ionization degree $\chi_\text{CZ} \sim 10^{-6}$) on the bow shock to the value close to ion number density on the contact discontinuity ($n_{i,c} \lesssim 0.12^{+0.12}_{-0.07}~\left( 0.33^{+0.25}_{-0.17} \right)\text{cm}^{-3}$). Additional ionization afterwards in the tail can increase the luminosity in \Ha{} to the observed value. As an independent confirmation, our scenario is in agreement with the difference between predicted and observed dispersion measure of PSR J2225+6535 ($\Delta DM\approx 23 ~\text{cm}^{-3}\text{pc} <  \Delta DM_{max} =  28 ~\text{cm}^{-3}\text{pc}$) and with observed rotation measure ($RM = -14.7 \pm 0.7~\text{rad}~\text{m}^{-2}$, $|RM| < RM_{max} \approx 80 ~\text{rad}~\text{m}^{-2}$).

Both these conclusions are important contributions to the models of particle acceleration by pulsars, and, independently, to the models of ISM. 

The problem of oblique collisionless shocks in ISM with a finite downstream length requires a consideration. Neutral ISM gas passing the bow shock undisturbed has much larger mean free path than the ionized component. Here we only tentatively grasped how the latter collisional medium arises from (and exists inside) the former collisionless one. There is no solid answer on how this large-scale delayed ionization modify the structure of the flow, how it affects the thickness of the shocked ISM zone and how exactly additional ISM ionization happens in the zone, which in ideal hydrodynamic approximation should be dominated by pulsar wind. 

In this work (see Section~\ref{subsec:ionization_structure}) we assume the number density on the contact discontinuity to be equal to that of ISM fraction being ionized. This does not affect the estimates of the bow shock dynamics (see \ref{sec:Bism_bowshock_config}) and suits our purpose of assessing the ionization structure within the precision of the order of the magnitude, but the question what number density of ions is the on contact discontinuity is still present. Also here we presume a perpendicularity of the pulsar velocity and the kinetic jet, which seems to be a fortunate coincidence. Differences in this angle across the nebulae \citep{2024ApJ...976....4D} must be incorporated in the model in order to expand on this work. 

The structure of bow shock considered here might be self-regulated: with an increase of the ISM number density the stand-off distance must become shorter. This in turn shortens the scale at which ionization occurs, making the fraction of ISM gas interacting with pulsar wind less, which can raise stand-off distance again to some stationary value. Collisionless shocks were modeled in regards to supernova remnants, with accounting of Ly$_\alpha$ trapping with Monte-Carlo and partial electron-proton equilibration \citep{2001ApJ...547..995G}, and also were considered analytically \citep{2007ApJ...654..923H}. The Guitar nebula requires comprehensive kinetic simulations in the future

Let us make estimates on the population of relativistic particles in the filament of Guitar nebula. The angular length of the filament is $l_f = 2.5$'. Flux in 0.5 -- 7.0~keV is $F_f = 5.49 \times 10^{-14}~\dfrac{\text{erg}}{\text{s cm}^2}$ \citep[we summed up fluxes from ``inner'', ``middle'', ``outer'', ``counter-filament'' and ``remnant'' parts from ][]{2022ApJ...939...70D}. The luminosity in the band is:
\begin{equation}
L_f = 4 \pi D^{2} F_{f} = 4.5 \times 10^{30}~\text{erg / s}~\left( \frac{D}{0.83~\text{kpc}} \right)^2.
\end{equation}

The synchrotron emission power of the single particle with Lorentz factor $\gamma_\text{em}$ is:
\begin{equation}
P_\text{syn} = \frac{4}{3} \gamma_\text{em}^{2} \sigma_{T} c u_{B},
\end{equation}
where $\sigma_T$ is Thomson cross section, $u_B = \dfrac{B_\text{ISM}^2}{8 \pi}$ is the magnetic energy density in the filament. Then the number of emitting particles is:
\begin{equation}
N_\text{em} = \frac{L_{f}}{P_\text{syn}}.
\end{equation}

With continuous injection of particles into the filament, its' power in emitting particles with respect to pulsar spin-down power is:
\begin{equation}
        \frac{P_{f,\text{em}}}{\dot{E}} = \frac{N_\text{em} \gamma_\text{em} c^{2} m_{e}}{\dot{E} \tau}, 
\end{equation}
where $\tau$ is a lifetime of the particles in the filament. It is between a cooling time ($\tau_\text{cool}$, in that case particles stay in the filament until they radiate their energy) and a flyover time ($\tau_\text{flyover}$, in that case particles propagate freely along the filament with the speed of light). The times are:
\begin{align} 
\tau_\text{cool} &= \frac{\gamma_\text{em} m_{e} c^{2}}{P_\text{syn}} = 25~\text{yr}~\gamma_{\text{em},8}^{-1} B_{\text{ISM},-4}^{-2}, \label{eq:t_cool} \\
\tau_\text{flyover} &= \frac{D l_{f}}{c} = 2.0~\text{yr}~\left( \frac{D}{0.83~\text{kpc}} \right) \left( \frac{l_f}{2.5'} \right).
\end{align} 
Thus, 
\begin{align}
        \frac{P_{f,\text{em}} \left( \tau = \tau_\text{cool} \right)}{\dot{E}} &= \frac{L_f}{\dot{E}} = 0.0045~\left( \frac{D}{0.83~\text{kpc}} \right)^2 \dot{E}_{33}^{-1}, \\ 
        \frac{P_{f,\text{em}} \left( \tau = \tau_\text{flyover} \right)}{\dot{E}} &= \frac{24 \pi^{2} m_{e} c^{2}}{\sigma_{T}} \frac{D F_{f}}{B_\text{ISM}^{2} \dot{E} \gamma_\text{em} l_{f}} =  \notag \\
        &= 0.056 \left( \frac{D}{0.83~\text{kpc}} \right) B_{\text{ISM},-4}^{-2} \dot{E}_{33}^{-1} \gamma_{\text{em},8}^{-1} \left( \frac{l_f}{2.5'} \right)^{-1}. \label{eq:P_f_t_cool} 
\end{align}

The estimates depend on the spin-down of the pulsar, they are 0.002 -- 0.004 in case of cooling and 0.02 -- 0.05 in case of flyover for the plausible range (see~\ref{subsec:spindown_estimate}). Here we must mention that the filament spectrum is wider than the considered band, so the estimates are only lower limits on the filament power. Moreover, average Lorentz factor of the particles in the filament is unknown. In Equations~\ref{eq:t_cool} and~\ref{eq:P_f_t_cool} we assumed $\gamma_\text{em} \sim 10^8$. Keeping that in mind, we conclude that the presented here scenario has a much softer energy constraints than the one with magnetic field generated with particles (see \ref{sec:mag_self_amp}). While the latter may occur in some nebulae, magnetic field around Guitar must be fossil.

The requirement on the number of the emitting particles is even more relaxed, if we assume the bulk Lorentz factor of the wind $\gamma_w = 10^6$. This is a speculative value, the lower cut-off of the particle spectrum is expected to be $\gtrsim 10^4$ for at least some of the filaments \citep{2025A&A...693A.192P}. The number of emitting particles with respect to all pulsar wind particles ($N_\text{tot}$) is then:
\begin{equation}
    \frac{N_\text{em}}{N_\text{tot}} = \frac{\gamma_w}{\gamma_\text{em}} \frac{P_{f,\text{em}}}{\dot{E}} = 10^{-2} \gamma_{w,6} \gamma_{\text{em},8}^{-1} \frac{P_{f,\text{em}}}{\dot{E}}.
\end{equation}
This is $[2; 4] \times 10^{-5}$ in case of cooling and $[2; 5] \times 10^{-4}$ in case of flyover for the plausible range of spin-down power.

In regards to extreme acceleration by pulsars and PWNe, they were proposed to be sources of very-high-energy (VHE; $\gtrsim 100$ GeV) gamma-rays (both extended halos and point sources). Examples include 1LHAASO J1740+0948u with spectrum extending up to 300~TeV, which is offset from the middle-aged pulsar PSR J1740+1000, the possible source of particles, in the direction of the tail of its X-ray nebula \citep{2025arXiv250215447C, 2026ApJ...997..341G}. Highly aged millisecond pulsar (MSP) J0218+4232 can accelerate particles to no more than $\sim$1~PeV ($10^{15}$ eV) and is the only known potential source of them for ``Peanut'' nebula, which was detected by LHAASO with photon energies up to 760~TeV and does not fit into halo templates \citep{2025arXiv251006786C}. If this association is confirmed, not only would ``Peanut'' have a huge size (250~pc $\times$ 25~pc), but the particle acceleration efficiency in terms of maximum possible energy would be extreme. The most striking example of extreme acceleration is PSR J1849-0001 ($\dot{E} \sim 10^{37}$~erg/s), known to power its' PWN visible in X-rays and VHE \citep{2026arXiv260315537T, 2026arXiv260319721A,2023A&A...672A.103H}. The highest energy photon registered from it is 2.3~PeV, which suggest acceleration efficiency close to 1 and can shed light on the origin of the particles more energetic than the cosmic ray ``knee''. Our research adds PSR J2225+6535 (with $\eta_\text{acc}\gtrsim 3/4$) to the list of the sources with extremely efficient acceleration and suggests that the latter may be a usual property of pulsars.

The escape of energetic particles from PWNe can occur in the form other than kinetic jets, namely {\it pulsar halos}. They are extended VHE emissions surrounding middle-aged pulsars ($\tau \sim 10^5$~yr). These structures, first unambiguously identified by the High-Altitude Water Cherenkov (HAWC) observatory around the Geminga and Monogem pulsars \citep{Abeysekara2017}, exhibit gamma-ray emission extending up to PeV energies \citep{Cao2021}. Their discovery challenges conventional models of particle acceleration and diffusion in pulsar wind nebulae (PWNe), while offering new insights into the origin of cosmic-ray (CR) electrons and positrons ($e^\pm$). The question on the condition defining the form of energetic particle escape is of a great interest. Today it is unknown, what governs whether PWN will form a kinetic jet or a halo. They might occur to be different counterparts of a single population of particles, firstly highlighting the local magnetic field in X-ray and then forming diffuse $\gamma$-ray emission. The building of a unified model of high energy manifestations of PWNe is up to future research.

In relation to the structure of ISM, the transition to the radiative phase in supernova remnants has long been hypothesized to lead to the   catastrophic collapse of the postshock gas, and  formation of a thin, dense shell \citep[e.g.][]{1981MNRAS.195.1011F, 1998ApJ...500..342B}. It is expected that a shell would then be highly magnetized  (we are not aware of any relevant  MHD calculations). We suggest that {\it Guitar} is currently traversing such a shell. The presence of magnetic field  may also resolve the puzzle: the cooled shells are expected to be RT unstable \citep{1983ApJ...274..152V,1993ApJ...407..207M}. We suggest that the compressed magnetic field  suppresses RT instabilities.

\section*{Acknowledgements}
This research was supported by the  BASIS foundation grant \#24-1-2-25-1. We thank Prof. Adam Deller for providing bootstrap astrometric solutions for PSR J2225+6535 and Prof. Roger Romani for the explanatory remarks on the measurement of {\it Guitar}  nebula bow shock stand-off distance. We appreciate N.N. Chugai and D.Z. Wiebe for useful discussion. We acknowledge using python packages sympy \citep{10.7717/peerj-cs.103}, numpy \citep{harris2020array}, astropy \citep{2022ApJ...935..167A}, scipy \citep{2020SciPy-NMeth}, matplotlib \citep{Hunter:2007}, tueplots\footnote{URL: https://github.com/pnkraemer/tueplots}, corner \citep{corner} and arviz \citep{arviz_2019}.

\appendix

\section{Reconstruction of parameters\label{sec:reconstruction}}

The main source of information about pulsar velocity is its proper motion ($\mu$) and the distance to the pulsar ($D$), which in turn strongly affects the bow shock stand-off distance. They were measured using VLBI observations for PSR J2225+6535 and listed in PSRPI catalogue \citep{2019ApJ...875..100D}. The measurement is done via a bootstrap algorithm, which provides the solution as a sample of probability distribution for astrometric quantities (initially 100000 samples). It also contains the celestial coordinates of the source. The distribution is strongly non-gaussian and multi-modal, so we propagate it directly in the form of samples, as provided by Prof. Adam Deller. We discard the samples outside the range $D \in [0.3;~3]$~kpc.  

Another important parameter is the inclination of the nebula ($i$), i.e. the angle between the nebula symmetry axis (and, in turn, pulsar velocity with respect to its' local medium for the axisymmetric wind) and the picture plane. \cite{2024ApJ...975L..31O} measured it for {\it Guitar}  nebula to be in range $-11^o \lesssim i \lesssim -7^o$ by fitting the velocities of blue-shifted and red-shifted broad components of \Ha\ line, which originate from approaching and receding sides of the bow shock in the tail region, with the equation derived from the one for tangent velocity from \cite{1996ApJ...459L..31W}. For simplicity here we assume that the measured quantity is normally distributed. To use it with astrometric data, which is in the form of a sampled distribution, we randomly generate samples from the assumed normal distribution of $i = (9 \pm 2)^o$.

\subsection{Pulsar velocity with respect to ISM\label{subsec:velocity}}

\cite{2019ApJ...875..100D} list the solutions with respect to the Sun and here we are interested in the velocity with respect to pulsar's local ISM. Therefore, the components of the transversal velocity ($V_\text{NS, ISM}^\text{transversal}$) are:
\begin{equation}
        \begin{bmatrix}
                V_{\text{NS, ISM},~l} \\
                V_{\text{NS, ISM},~b} 
        \end{bmatrix} = D \begin{bmatrix}
                \mu_l \cos{b} \\
                \mu_b
        \end{bmatrix} - \begin{bmatrix}
                V_{\text{ISM}, \odot, l} \\
                V_{\text{ISM}, \odot, b}
        \end{bmatrix}.
\end{equation}\label{eq:velocity_from_pm}
Here $l$ and $b$ are the galactic latitude and the galactic longitude. The projections to their tangent vectors are denoted by the corresponding indices. In order to perform the correction, we find the components of pulsar's local ISM velocity with respect to the Sun ($V_{\text{ISM}, \odot, l} $ and $V_{\text{ISM}, \odot, b}$) \citep{2025ApJ...980...80S}. It is implied that the interstellar matter moves with approximately the same velocity as the stellar population, described by Oort constants \citep{2017MNRAS.468L..63B} and the velocity of the Sun \citep{2010MNRAS.403.1829S}. Here we do not propagate the systematic uncertainties, since their values are $\sim$few~km/s, which is much less than the uncertainty of the pulsar velocity.

Since the inclination is determined by the method, which traces the interaction of pulsar wind with local ISM \citep{2024ApJ...975L..31O}, we infer:
\begin{equation}
        V_\text{NS} \equiv V_\text{NS, ISM}  = \frac{V_\text{NS, ISM}^\text{transversal}}{\cos{i}}.
\end{equation}

\begin{figure*}
        \centering
        \includegraphics[width=\textwidth]{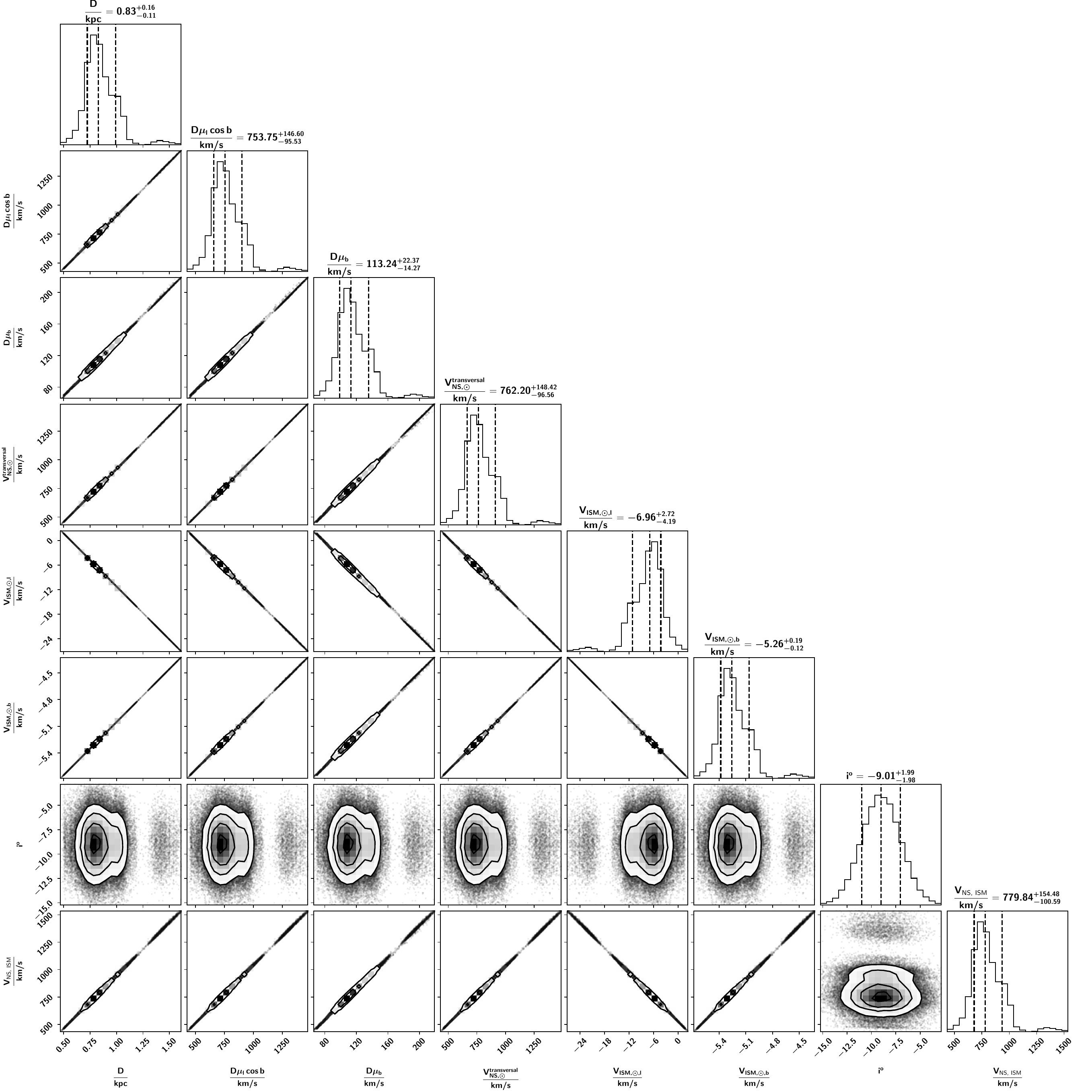}        
        \caption{Corner plot of pulsar velocity and the quantities used for the reconstruction of it. Namely, from left to right: distance to the pulsar ($D$), transversal velocity of the pulsar with respect to the Sun projected to tangent vectors of galactic coordinate system ($D \mu_l \cos{b}$ and $D \mu_b$) and its value ($V_{\textbf{NS}, \odot}^\textbf{transversal}$), velocity of pulsar's local ISM projected to tangent vectors of galactic coordinate system ($V_{\textbf{ISM}, \odot, l}$ and $V_{\textbf{ISM}, \odot, b}$), inclination of {\it Guitar}  nebula ($i$) and the velocity of the pulsar with respect to its local ISM ($V_\text{NS, ISM}$)}
        \label{fig:reconstruction_velocity}
\end{figure*}

The results of pulsar velocity reconstruction are listed in Figure~\ref{fig:reconstruction_velocity}. It is presented mainly to show the propagation of distributions of quantities through the equations. In cases where uncertainty of one parameter governs the result, there is a strong linear trend in the figure. Conversely, on the panels where uncorrelated input parameters are presented or the result is influenced by the uncertainties of other parameters in case of propagation, 2D distribution is close to symmetric one.

Transversal velocity of the pulsar with respect to the Sun is  $V_{\text{NS}, \odot}^\text{transversal} = 762^{+148}_{-96}$~km/s, which is very close to the value $765^{+158}_{-93}$, reported in PSRPI \citep{2019ApJ...875..100D}. The overall correction for galactic rotation and projection effect is 18~km/s, which is small with respect to resulting velocity $V_\text{NS} = 780^{+154}_{-100}$~km/s and its uncertainty. Nevertheless, since the direct measurement of pulsar radial velocity is not possible, here we have one of the few cases where the full 3D reconstruction of pulsar velocity is performed \citep[also see][]{2022ApJ...930..101R}.

\subsection{Stand-off distance}

The observational data is an angular projected stand-off distance of Guitar nebula bow shock ($\rho_0$) \citep{2022ApJ...939...70D}. We use the results for year 2006, since they are the closest to VLBI observations (year 2011) \citep{2019ApJ...875..100D} and X-ray observations \citep{2007A&A...467.1209H, 2012ApJ...747...74H}. Hubble Space Telescope \Ha\ high angular resolution observations can be fitted with two models: ``thin shock'' model (which yields $\rho_0=97^{+4}_{-4}$~mas), which describes momentum-balanced bow shock \citep{1996ApJ...459L..31W, 2000ApJ...532..400W}, and ``thick shock model'' (which yields $\rho_0=93^{+3}_{-3}$~mas), which empirically describes the observed widening of the shock \citep{2022ApJ...939...70D}. Noting that the results of these models are well within the margin of error, here we employ the former, since our analytical model is also built in momentum-balanced shock formalism. We generate random values for projected angular standoff distance following the provided histogram and add them to the sample from distribution of other quantities for further analysis.

This allows one to reconstruct the projected stand-off distance $r_0$ from the angular distance $\rho_0$ and the distance to the pulsar $D$:
\begin{equation}\label{eq:projected_standoff}
r_{0} = D \rho_{0}.
\end{equation}

The dependence of actual stand-off distance ($r_s$) from projected one is not straightforward. The position of the limb of the observed nebula given the inclination ($i$) is determined by the geometry of the shock, namely it is in the place where the emitting layer is aligned with the line of sight \citep{2025PASA...42...79N, 2020MNRAS.497.2605B}. Thus, if the shape of the contact discontinuity in polar coordinates $(\theta, r)$ is determined by function $r = r_s f(\theta)$, we have:  
\begin{equation}\label{eq:projected_standoff_vs_actual}
        r_0 = r_s f(|i|).
\end{equation}

There was an argument that the shape of the bow shock is weakly influenced by the external magnetic field and the anisotropy of the pulsar wind, the main influence being a widening of the structure by 20 -- 30\% and an introduction of a slight asymmetry \citep{2019MNRAS.484.4760B}. It may explain the good quantitative fit to Guitar nebula head \citep{2022ApJ...939...70D} and qualitative similarities to PSR J2030+4415 \citep{2020ApJ...896L...7D} using the formalism for a momentum balanced shock \citep{1996ApJ...459L..31W, 2000ApJ...532..400W}:
\begin{equation}\label{eq:distances}
        f(\theta) = \csc{\left(\theta \right)} \sqrt{3 \left(1 -  \theta \cot{\left(\theta \right)} \right)}.
\end{equation}

\begin{figure*}
        \centering
        \includegraphics[width=\textwidth]{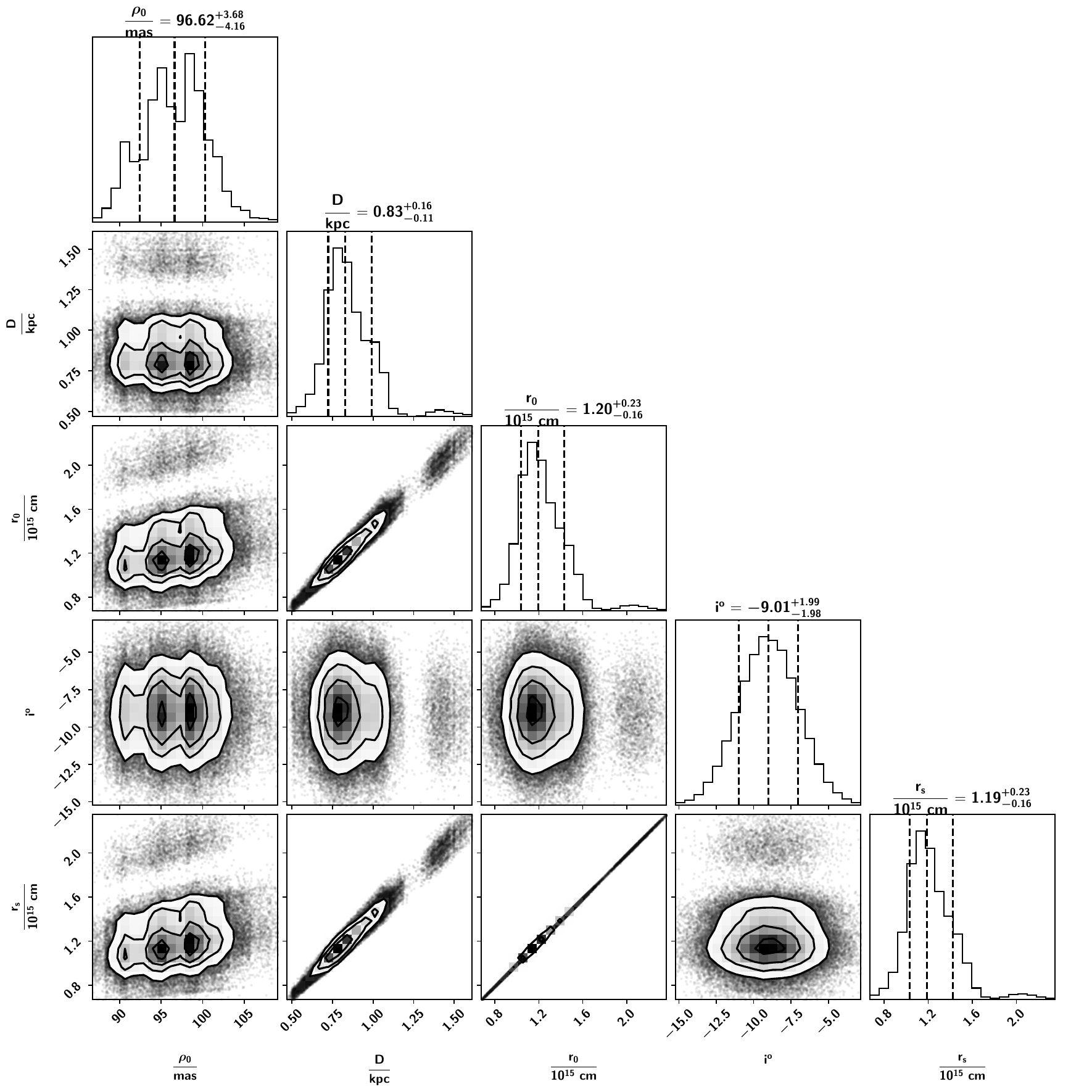}        
        \caption{Corner plot of Guitar bow shock standoff distance and the quantities used for the reconstruction of it. Namely, from left to right: angular projected stand-off distance ($\rho_0$), distance to the pulsar ($D$), projected stand-off distance ($r_0$), inclination of Guitar nebula ($i$) and the actual reconstructed stand-off distance ($r_s$)}
        \label{fig:reconstruction_standoff}
\end{figure*}

We use Equations~(\ref{eq:projected_standoff}), (\ref{eq:projected_standoff_vs_actual}) and (\ref{eq:distances}) to reconstruct the actual deprojected stand-off distance $r_s$. The result is shown in Figure~\ref{fig:reconstruction_standoff}. We note that the correction for the effect of inclination  ($\sim 10^{13}$~cm) is much less than the uncertainty of the stand-off distance and its value ($r_s = 1.19^{+0.23}_{-0.16} \times 10^{15}$~cm).

\section{Spin-down power of PSR J2225+6535\label{subsec:spindown_estimate}}

The timing of the pulsar was measured over the course of more than 1000 days \citep{2006A&A...457..611J}, the values of frequency and its derivatives are $\nu = 1.465113851(1 \pm 2)~\text{s}^{-1}$, $\dot{\nu} = -20.77(4 \pm 3) \times 10^{-15}~\text{s}^{-2}$, and $\ddot{\nu} = -0.5(5 \pm 3) \times 10^{-24}~\text{s}^{-3}$. 

We assess  $\dot{E} =  - I \omega  \dot{\omega}  = - 4 \pi^2 I \nu  \dot{\nu} $, where $I$ is the moment of inertia of the neutron star, $\omega$ is the angular velocity of its rotation and $\nu$ is the frequency of the pulses. Using the canonical value of $I = 1 \times 10^{45}~\text{g}~\text{cm}^2$, we assess the spin-down power as $\dot{E} = 1.2 \times 10^{33}$~erg/s, which is the same as in Australia Telescope National Facility catalogue \citep[ATNF, ][]{2005AJ....129.1993M}.

The moment of inertia is not known accurately, because it depends on the equation of state (EOS) of the neutron star. Nevertheless we can estimate it, analyzing the existing EOSs in $I/M^{3/2} - M$ coordinates \citep{2005ApJ...629..979L, 2007PhR...442..109L}. The heaviest known neutron star has a mass of $M = 2.35 \pm 0.17 M_\odot$ and the estimated limitations on the maximum possible mass of a neutron star are $> 2.19 M_\odot (2.09 M_\odot)$ at $1 \sigma$ ($3 \sigma$) confidence \citep{2022ApJ...934L..17R}. We select EOSs permitting this estimate (at $3 \sigma$ level), except for extremely stiff MS0 \citep{1996NuPhA.606..508M}, which struggles to reproduce existing observational constraints on mass and radius \citep{2002ApJ...576L.145W, 2014ApJ...796L...3G}. WFF1 \citep{1988PhRvC..38.1010W} yields the lowest value of $I / M^{3/2} = 35~\text{km}^{1/2} / M_\odot^{1/2}$ and AP3 \citep{1997PhRvC..56.2261A} yields the highest one of $45~\text{km}^{1/2} / M_\odot^{1/2}$ for most values of $M$. 

The mass distribution of slow pulsars (i.e., the small spin period pulsars and neutron stars with high-mass companions, which are likely to be near their birth masses) has the mean value of $\langle M \rangle = 1.49 M_\odot$ and dispersion of $\sigma_M = 0.19 M_\odot$ \citep{2016ARA&A..54..401O}. We infer $I(\langle M \rangle - \sigma_M) = 1.03 \times 10^{45}~\text{g}~\text{cm}^2$ with WFF1 and $I(\langle M \rangle + \sigma_M) = 1.95 \times 10^{45}~\text{g}~\text{cm}^2$ with AP3, which corresponds to spin-down power in the range $\dot{E}_\text{min} \left(= 1.2 \times 10^{33}~\text{erg/s}\right) \lesssim \dot{E} \lesssim \dot{E}_\text{max} \left(=2.3 \times 10^{33}~\text{erg/s}\right)$. We perform calculations with both of these edge cases. We write the values obtained with $\dot{E} = \dot{E}_\text{min}$ without brackets and the ones with $\dot{E} = \dot{E}_\text{max}$ with brackets.

\section{On amplification of the ISM magnetic field  by escaping particles\label{sec:mag_self_amp}}

There is an alternative scenario of strong magnetic field formation \citep{2024A&A...684L...1O} based on the idea of magnetic field amplification due to  Bell's instability \citep{2004MNRAS.353..550B}. In this scenario, the current of the escaped particles generates the magnetic field of the filament. 
Let's investigate the very basic energetic limitation of such scenario. For that, we need to estimate the X-ray filament volume growth rate.  
The angular length of the filament is $l_f = 2.5$'. X-ray luminosity profile averaged over the length is described with a Gaussian profile with $\sigma_{2000} = 3.05^{+1.23}_{-1.17}$'', $\dot{\sigma}_{2000} = 0.25^{+0.09}_{-0.07}$''/yr. The profile moves with proper motion $\mu = 0.15^{+0.07}_{-0.09}$''/yr \citep{2022ApJ...939...70D}. XMM-Newton observations, from which we derive our estimates, were taken in July 2009 \citep{2012ApJ...747...74H}, for that epoch we estimate the width of the filament as $\sigma = \sigma_{2000} + \dot{\sigma}_{2000} \times 9.5~\text{yr} = 5.4$''. If the filament is considered a cylinder with averaged width, it sweeps a volume of space, in which magnetic field must be generated, at a rate:
\begin{align}
        \dot{V} &= 2 \sigma l_f \mu D^3 = 1.5 \times 10^{43}~\frac{\text{cm}^3}{\text{s}} \times \\ \notag
                &\times \left( \frac{\sigma}{5.4''} \right) \left( \frac{l_f}{2.5'} \right) \left( \frac{\mu}{0.15''/\text{yr}} \right) \left( \frac{D}{0.83~\text{kpc}} \right)^3.
\end{align}
The power required for generation of the magnetic field in this volume is:
\begin{equation}
        P_f > \frac{B_{min}^2}{8 \pi} \dot{V} = 4 \times 10^{33}~B_{-4}^2 \dot{V}_{43}~\frac{\text{erg}}{\text{s}}.
\end{equation}

In our model the minimum possible magnetic field (see Equation~(\ref{eq:B_ism_Xray})) is determined only by the pulsar energy budget ($\dot{E}$) and the filament spectrum ($E_\text{cutoff}$), and its strength for different estimates on spin-down power (see Table~\ref{tab:model_parameters_bootstrap}) \footnote{ The magnetic field value  $78~\mu$G obtained in \citep{2024A&A...684L...1O} is inside our limitation for various spin-down power of pulsar  (see Table~\ref{tab:model_parameters_bootstrap}). }. Thus, with the most optimistic estimates we get $P_f  = 11\times10^{33}~\text{erg/s}$ for $\dot{E}_{min}$ or  $3\times10^{33}~\text{erg/s}$ for $\dot{E}_{max}$, so $P_f$ exceeds the total pulsar spin-down power $\dot{E}$ from 9 to 1.4 times respectively.  

Another problem is the dynamically unbalanced magnetic field, whose energy density is comparable with pulsar wind pressure on the bow shock of the nebula. In the magnetic field generation scenario, this should create a shock front in ISM visible in H$_\alpha$ line, which is not detected so far.

\section{Ionization state evolution between the bow shock apex and contact discontinuity \label{sec:Ionization}}
Considering that in the apex the shock velocity equals the neutron star velocity, we compute post-shock electron temperature there \citep{2002ApJ...572..888G}:
\begin{equation}\label{eq:Te_noneq}
        T_e = \frac{3 }{16} \frac{m_{p} V_\text{NS}^{2}}{k_{B}} \left(\mu_{i} f_\text{eq} + \frac{m_{e}}{m_{p}} \left(1 - f_\text{eq}\right)\right),
\end{equation}
where $\mu_{i} = 0.6$ is the molecular mass of ionized gas and $f_\text{eq}$ is equilibration parameter, varying from 0 for the conservation of momentum ($T_e = \frac{m_e}{m_p} T_p$; here $T_p$ is proton temperature) to 1 for full Coulomb equilibration ($T_e = T_p$). Following \cite{2007ApJ...654..923H}, we considered $f_\text{eq}$ of 0.35 (better approximates hydrogen lines wide-to-narrow components ratio for SNRs) and 1, which results in $T_e$ of $(2.9^{+1.3}_{-0.7}) \times 10^6$~K and $(8.3^{+3.6}_{-2.0}) \times 10^6$~K. The difference is high, but for a few million kelvin the ionization coefficient for hydrogen ($C$) is weakly influenced by the temperature \citep{2007A&A...466..771D}, and interpolation yields $(2.82^{+0.06}_{-0.13}) \times 10^{-8}~\text{cm}^3 s^{-1}$ and $(2.32^{+0.16}_{-0.22}) \times 10^{-8}~\text{cm}^3 s^{-1}$ respectively. 

The contact discontinuity is acted upon by ionized gas, while in ambient medium it is mostly neutral. Ionization begins when it passes the bow shock and continues until it reaches the contact discontinuity. Thus we infer the ionization timescale as flyover time of perturbed ISM: $t \approx 0.3 \frac{r_s}{V_\text{NS}} = 0.145^{+0.006}_{-0.006}$~yr.

Since $\frac{d n_e}{d t} \approx \frac{d n_{\text{H}^+}}{d t} = n_e n_\text{H} C$, where $n_\text{H}$ and $n_{\text{H}^+}$ are number densities of neutral and ionized hydrogen, as long as $n_e \ll n_\text{H}$, for timescale $t$ and $n_{e, i} = n_e (0)$,
\begin{equation} 
n_{\text{H}^+} = n_e = n_{e, i} e^{n_\text{H} C t}.
\end{equation}
Thus, in our scenario for $n_{\text{H}^+} \approx 0.1~\text{cm}^{-3}$ being $n_{i,c}$ from Table~\ref{tab:model_parameters_bootstrap} and $n_\text{H} \approx 100~\text{cm}^{-3}$, we infer $n_{e, i} \approx 10^{-4}~\text{cm}^{-3}$. The pre-shock ionization degree is $\chi = \frac{n_{e, i}}{n_\text{H}} \sim 10^{-6}$. This is a typical value for relatively dense ISM regions and supports our scenario. We note that only a small fraction of the gas is being ionized between the shock and the contact discontinuity in this fast PWN with unusually small head.

\bibliographystyle{elsarticle-harv} 
\bibliography{ismpw.bib}

\end{document}